\documentclass[%
 reprint,
 superscriptaddress,
 amsmath,amssymb,
 aps,
prl
floatfix,
]{revtex4-2}
\usepackage{graphicx}
\usepackage{dcolumn}
\usepackage{bm}
\usepackage[utf8]{inputenc}
\usepackage[T1]{fontenc}
\usepackage{mathptmx}
\usepackage{etoolbox}
\usepackage{calc}
\usepackage{comment}
\usepackage{graphicx}
\usepackage[shortlabels]{enumitem}
\usepackage{amsmath}
\usepackage{amsfonts}
\usepackage{amssymb}
\usepackage{float}
\usepackage{xcolor}
\usepackage{spalign}
\usepackage{natbib}
\usepackage{systeme}
\usepackage{units}
\usepackage{physics}
\DeclareMathOperator\FWHM{FWHM}
\newcommand{\angstrom}{\textup{\AA}}
\usepackage{epsfig}
\usepackage{color} 

\begin{document}
\DeclareGraphicsExtensions{.eps,.jpg,.png}
\input epsf
\title{Probing enhanced electron-phonon coupling in graphene by infrared resonance Raman spectroscopy}
\author{Tommaso Venanzi}
\affiliation{Department of Physics, Sapienza University of Rome, Piazzale Aldo Moro 5, I-00185 Rome, Italy}
\author{Lorenzo Graziotto}
\affiliation{Department of Physics, Sapienza University of Rome, Piazzale Aldo Moro 5, I-00185 Rome, Italy}
\author{Francesco Macheda}
\affiliation{Istituto Italiano di Tecnologia, Graphene Labs, Via Morego 30, I-16163 Genoa, Italy}
\author{Simone Sotgiu}
\affiliation{Department of Physics, Sapienza University of Rome, Piazzale Aldo Moro 5, I-00185 Rome, Italy}
\author{Taoufiq Ouaj}
\affiliation{JARA-FIT and 2nd Institute of Physics, RWTH Aachen University, 52074 Aachen, Germany}
\author{Elena Stellino}
\affiliation{Department of Physics and Geology, University of Perugia, via Alessandro Pascoli, Perugia, Italy}
\author{Claudia Fasolato}
\affiliation{Institute for Complex System, National Research Council (ISC-CNR), 00185 Rome, Italy}
\author{Paolo Postorino}
\affiliation{Department of Physics, Sapienza University of Rome, Piazzale Aldo Moro 5, I-00185 Rome, Italy}
\author{Vaidotas Mi\v{s}eikis}
\affiliation{Istituto Italiano di Tecnologia, Graphene Labs, Via Morego 30, I-16163 Genoa, Italy}
\affiliation{Istituto Italiano di Tecnologia, Center for Nanotechnology Innovation @NEST, Piazza San Silvestro, 12-56126 Pisa, Italy}
\author{Marvin Metzelaars}
 \affiliation{Institute of Inorganic Chemistry, RWTH Aachen University, 52074, Aachen, Germany}
 \author{Paul Kögerler}
 \affiliation{Institute of Inorganic Chemistry, RWTH Aachen University, 52074, Aachen, Germany}
\author{Bernd Beschoten}
 \affiliation{JARA-FIT and 2nd Institute of Physics, RWTH Aachen University, 52074 Aachen, Germany}
\author{Camilla Coletti}
\affiliation{Istituto Italiano di Tecnologia, Graphene Labs, Via Morego 30, I-16163 Genoa, Italy}
\affiliation{Istituto Italiano di Tecnologia, Center for Nanotechnology Innovation @NEST, Piazza San Silvestro, 12-56126 Pisa, Italy}
\author{Stefano Roddaro}
\affiliation{Department of Physics, University of Pisa, Largo B. Pontecorvo 3, I-56127 Pisa, Italy}
\author{Matteo Calandra}
\affiliation{Department of Physics, University of Trento, Via Sommarive 14, 38123 Povo, Italy}
\author{Michele Ortolani}
\affiliation{Department of Physics, Sapienza University of Rome, Piazzale Aldo Moro 5, I-00185 Rome, Italy}
\author{Christoph Stampfer}
\affiliation{JARA-FIT and 2nd Institute of Physics, RWTH Aachen University, 52074 Aachen, Germany}
\author{Francesco Mauri}
\affiliation{Department of Physics, Sapienza University of Rome, Piazzale Aldo Moro 5, I-00185 Rome, Italy}
\affiliation{Istituto Italiano di Tecnologia, Graphene Labs, Via Morego 30, I-16163 Genoa, Italy}%
\author{Leonetta Baldassarre}
\affiliation{Department of Physics, Sapienza University of Rome, Piazzale Aldo Moro 5, I-00185 Rome, Italy}

\begin{abstract}
 
We report on resonance Raman spectroscopy measurements with excitation photon energy down to 1.16 eV on graphene, to study how low-energy carriers interact with lattice vibrations. Thanks to the excitation energy close to the Dirac point at $\mathbf{K}$, we unveil a giant increase of the intensity ratio between the double-resonant 2D and 2D$^\prime$ peaks with respect to that measured in graphite. Comparing with fully \textit{ab initio} theoretical calculations, we conclude that the observation is explained by an enhanced, momentum-dependent coupling between electrons and Brillouin zone-boundary optical phonons. This finding applies to two dimensional Dirac systems and has important consequences for the modeling of transport in graphene devices operating at room temperature.
\end{abstract}
\pacs{}
\maketitle

The presence of Dirac fermion excitations in the low-energy electronic structure is typical of graphene as well as topological insulators \cite{wehling2014dirac}. They share universal behaviors such as suppressed backscattering and similar transport properties and optical conductivity \cite{sarma2011electronic}. The competing interactions in these systems, such as electron-phonon and electron-electron correlations, can impact the low-energy states leading to Fermi velocity renormalization \cite{elias2011dirac} and the generation of mass-gaps \cite{feldman2009broken}, as well as resulting in superconducting states \cite{uchoa2007superconducting}. In graphene, the paradigmatic example of Dirac materials, the electron-phonon coupling (EPC) determines several features of both the phonon and electron properties such as (i) the room temperature (RT) resistivity, (ii) the behavior of the Kohn anomalies around $\mathbf{K}$ and $\boldsymbol{\Gamma}$~\cite{pisana2007breakdown, hasdeo2016fermi, mafra2009observation} and (iii) the kink found in ARPES experiments \cite{bostwick2007quasiparticle}. However, the strength of the EPC for low-energy carriers close to the Dirac point is not clear yet. It was shown theoretically, via simplified but general low-energy models, that the Coulomb interaction induces, in two dimensional materials, a strong enhancement of the EPC in the neighborhood of the Dirac point \cite{basko2008interplay}. An indirect evidence of this mechanism may be found in the underestimation of the resistivity of low-doping graphene  typical of density functional theory (DFT) first-principles calculations~\cite{PhysRev.136.B864,PhysRev.140.A1133,PhysRevB.90.125414,park2014electron,Macheda2020}, that underestimate the coupling with optical phonons because they neglect the Coulomb vertex corrections\cite{PhysRevB.90.125414}. The importance of dimensionality on such effects is remarked by the case of graphite, where it has been shown that the theoretical evaluation of the EPC close to the Dirac point is in agreement with the experiments \cite{lazzeri2008impact}.

Experimentally, estimates of the EPC can be obtained via the study of resonance Raman spectra~\cite{ferrari2013raman}. The Raman spectrum of graphene is understood in terms of first-order non-resonant scattering process with phonon momentum \textit{{\bf q}} = 0 (the G peak) and  higher-order processes, interpreted within the double resonance mechanism, that have momentum $\mathbf{q} \neq 0$ and are activated either via scattering with a phonon and a defect (D and D$^\prime$) or with two phonons of opposite momenta (2D and 2D$^\prime$)~\cite{ferrari2006raman, malard2009raman}.  The double resonant Raman processes involve electron and hole scattering within the Dirac cone or between neighboring cones via the emission of, respectively, zone-center ($\vb{q} \sim \boldsymbol{\Gamma}$) or zone-boundary ($\vb{q} \sim \mathrm{\textbf{K}}$) phonons. By changing the excitation laser energy ($\hbar\omega_L$), different regions of the electron and phonon dispersions of graphene ($\epsilon_{\mathbf{k}}$ and $\omega_{\mathbf{q}}$, where $\mathbf{k}$ and $\mathbf{q}$ are the electron and phonon momentum, respectively) can be optically probed thanks to the peculiar existence in graphene of a resonance condition at any $\omega_{L}$~\cite{mafra2007determination}. Therefore, any variation of the probed electron and phonon regions with $\hbar\omega_L$ is translated into a modification of the intensity, width, and position of the Raman peaks~\cite{venezuela2011theory}, whose analysis, in turn, yields direct information on the EPC. By moving $\hbar\omega_L$ in the infrared, one can thus assess the value of the EPC for low-energy carriers and derive general scaling behaviours valid for any two dimensional Dirac system.

Up to now, the EPC in graphene has been assumed constant in the vicinity of both the $\bf{K}$ and $\boldsymbol{\Gamma}$ points and its experimental estimates were obtained by comparison with graphite Kohn anomalies \cite{PhysRevB.80.085423}, by modeling the resistivity \cite{park2014electron}, and by looking at the intensity of the double resonant peaks \cite{ferrari2007raman}. The intensity of the double-resonant Raman peaks 2D and 2D$^\prime$ (i.e.\ the integrated area underneath these vibrational peaks, $A_{2D}$ and $A_{2D'}$) is a very sensitive tool since it is directly proportional to the fourth power of the EPC constant close to $\mathbf{K}$ and $\boldsymbol{\Gamma}$, respectively~\cite{basko2008theory,venezuela2011theory}. The ratio $A_{2D}/A_{2D'}$ is expected to depend non trivially on the phonon momentum, i.e. on the distance from the $\bf{K}$ point~\cite{basko2008interplay}; however, up to now, it has been evaluated experimentally at few $\hbar\omega_L$ for visible excitation energies only~\cite{ferrari2006raman}, too far from the $\bf{K}$ point to characterize such a behaviour.

The aim of this work is to address optically the electron-phonon coupling in graphene by means of resonance Raman spectroscopy measurements with excitation energies from 3.06 eV to $1.16\;$eV, thus exploring a resonant region of the electron and phonon dispersions close to the $\bf{K}$ point ($|\mathbf{k}-\mathbf{K}|\sim 0.07 \angstrom^{-1}$ and $|\mathbf{q}-\mathbf{K}|\sim 0.16 \angstrom^{-1}$, respectively). There, we find a huge enhancement of the $A_{2D}/A_{2D'}$ ratio in graphene that we interpret in terms of a strong dependence of the EPC upon the vicinity of $\mathbf{q}$ to the $\mathbf{K}$ point; by comparing measurements on graphene and graphite, we confirm that the enhancement of the EPC close to the  $\mathbf{K}$ point is strongly dependent on the dimensionality of the system. We further compare the experimental results to \textit{ab initio} calculations to quantitatively address the enhancement and to seamlessly reproduce the experimental trend of the 2D peak linewidth.  

\begin{figure}
\centering
\includegraphics[scale=0.46]{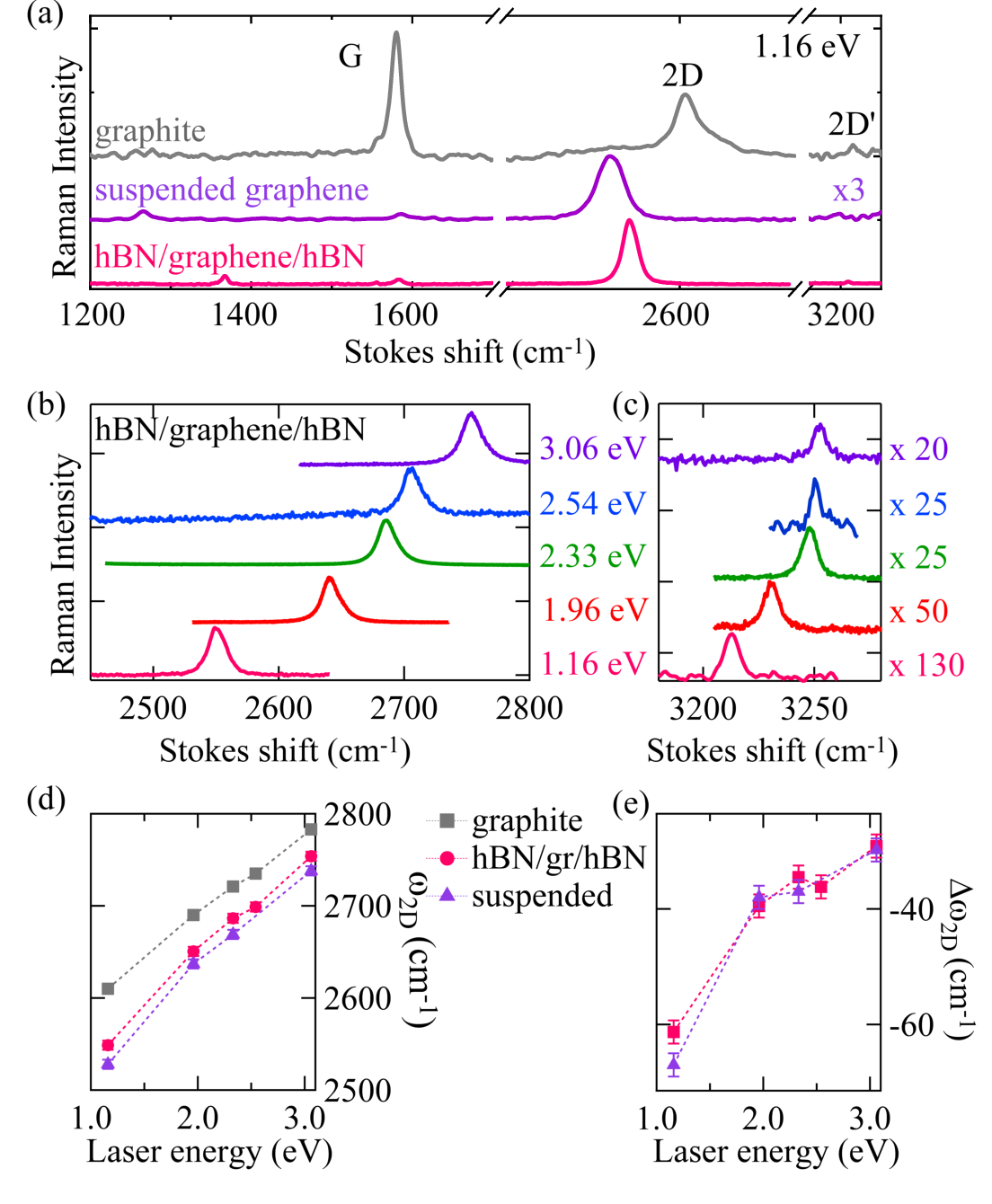}
\caption{(a) FT-Raman spectra of graphite (HOPG), hBN-encapsulated graphene (hBN/graphene/hBN), and suspended graphene. The G, 2D and 2D$^\prime$ peaks are indentified for all the three samples. (b) 2D and (c) 2D$'$ resonance Raman peaks collected at five different $\hbar\omega_L$  (indicated in figure). Data have been normalized to 2D peak intensity and different multiplying factors have been used to enhance the visibility of 2D$'$ peak. (d)  Dispersion of the 2D peak as a function of laser energy $\hbar\omega_L$; (e) difference between the 2D peak center frequency measured on graphene and graphite ($\Delta\omega_{2D}= {\omega_{2D}}^{graphene} - {\omega_{2D}}^{graphite} $), for both suspended and encapsulated samples. For clarity the data for suspended graphene were rigidly shifted upwards of 15 cm$^{-1}$. $\Delta\omega_{2D}$ is fairly constant for visible laser energies, and it steepens towards 1.16 eV. The values reported in the manuscript are derived from the analysis of single spectra,  acquired as the average of many repeated spectra with long integration time. Further details on the measurement and fitting procedure, and on the error bars can be found in Section 4 of SM.
}
\label{fig1}
\end{figure}

We have measured graphene flakes either exfoliated and encapsulated with PDMS stamps in hBN, or grown by CVD and transferred on a hole array on a Si$_3$N$_4$ membrane on Si~\cite{miseikis2017deterministic, giambra2021wafer,}. Highly oriented pyrolytic graphite (HOPG, from HQ graphene) was exfoliated to have a freshly cleaved surface to measure. Raman spectra have been collected with either dispersive setups at $\hbar\omega_L$ = 1.96, 2.33, 2.54 and 3.06 eV or with a Fourier-Transform one for $\hbar\omega_L$ = 1.16 eV (see SM for the experimental details).  
 
\begin{figure} [ht!]
\centering
\includegraphics[scale=0.52]{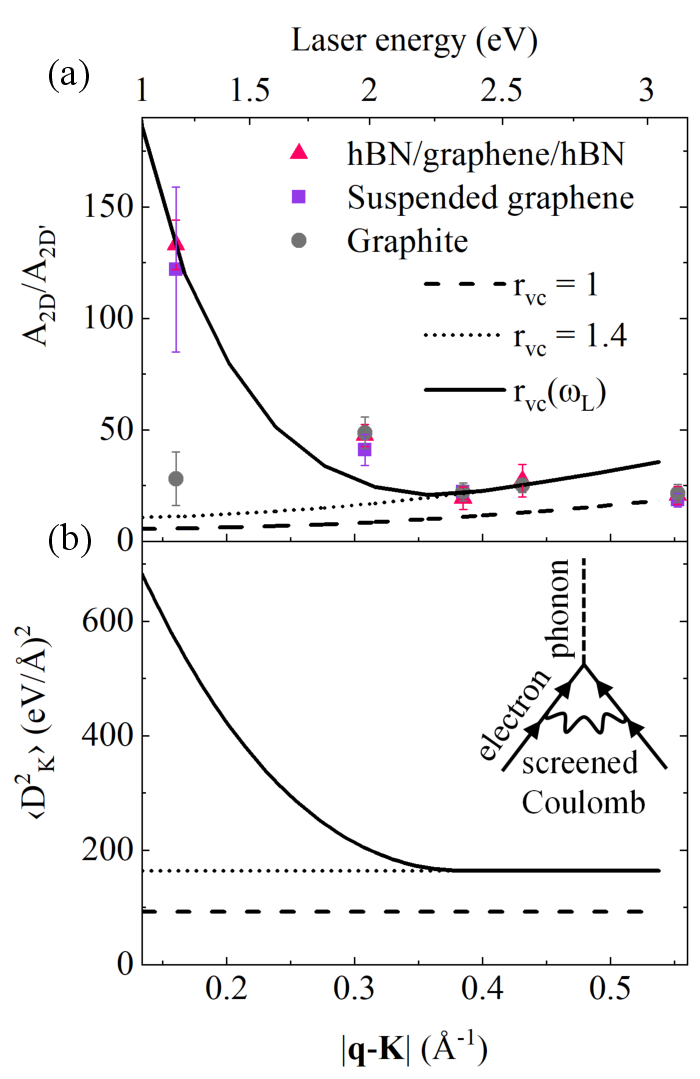}
\caption{(a) Ratio $A_{2D}/A_{2D'}$ as a function of the laser energy and of the position of the phonon resonance along the $\boldsymbol{\Gamma}-\mathbf{K}$ line. The experimental data points are compared to theoretical calculations rescaled with the square of $r_{vc}(\omega_L) \equiv\langle D_{\mathbf{K}}^2\rangle/2\langle D_{\boldsymbol{\Gamma}}^2\rangle$, where $r_{vc}=1$ (dashed black line), $r_{vc}=1.4$ (dotted black line), or $r_{vc}=r_{vc}(\omega_L)$, defined below in Eq.~(\ref{eq:r_cv}) (solid black line). The large experimental increase of the ratio value in the infrared indicates a strong increase of the EPC at the $\bf{K}$ point with respect to that at the $\boldsymbol{\Gamma}$ point. (b) Value of $\langle D_{\mathbf{K}}^2\rangle$ in the vicinity of \textbf{K} deduced from $r_{vc}(\omega_L)$ as explained in the text. Sketch: leading contribution to the vertex corrections in the EPC in graphene~\cite{basko2008interplay}.} 
\label{fig2} 
\end{figure}

In Fig.~\ref{fig1}(a) we report the Raman spectra for encapsulated and suspended graphene samples at $\hbar\omega_L = 1.16$ eV, comparing it to that measured on graphite. The 2D$^\prime$ peak is extremely weak, however above the noise level (see Fig.~\ref{fig1}(c)). We notice that the 2D peak of the suspended graphene is slightly blue-shifted due to tensile strain, as previously seen~\cite{berciaud2013intrinsic, shivaraman2009free, colangelo2018controlling}, and broader mainly due to a higher number of defects, as suggested by the presence of the D peak, that is instead negligible for both graphite and hBN encapsulated graphene. In Fig.~\ref{fig1}(b) we show the Raman 2D peak obtained for different $\hbar\omega_L$ from which the phonon dispersion shown in Fig.~\ref{fig1}(d) is deduced; here, we notice that the slope of the graphene dispersion appears to steepen with respect to the graphite one between $\hbar\omega_L $= 1.96 eV and 1.16 eV (better visible in Fig. \ref{fig1}(e)). This result is in qualitative accordance with a stronger enhancement of the EPC near the $\mathbf{K}$ point in graphene with respect to graphite, because of the lower dimensionality \cite{PhysRevLett.93.185503}. 

The intensity of the double-resonant Raman peaks of given phonons is the most sensitive marker to evaluate any modification of the EPC between equal resonant electronic bands as a function of $\mathbf{q}$ and $\mathbf{k}$; more precisely, such coupling is proportional to $D_{\mathbf{k}+\mathbf{q},\mathbf{k}}$ \cite{lazzeri2005electron}, which is defined as the variation of the band energy at $\mathbf{k}$ due to the atomic displacement induced by the phonon with momentum $\mathbf{q}$ commensurate with the lattice supercell ~\cite{lazzeri2008impact}. For our aims it is convenient to consider the quantity $\langle D_{\mathbf{K}/\boldsymbol{\Gamma}}^2 \rangle$, where $\langle\dots\rangle$ is the average over all the electronic momenta $\vb{k} \sim \mathrm{\textbf{K}}$ and phonon momenta $\vb{q} \sim \mathrm{\textbf{K}}$ ($\vb{q} \sim \boldsymbol{\Gamma}$) that satisfy the resonance condition for the 2D (2D$^\prime$) peak. In this notation, the intensity of the peaks is proportional to $\langle D_{\mathbf{K}/\boldsymbol{\Gamma}}^2\rangle^2$. On the other hand the widths and the spectral positions of the 2D and 2D' peaks provide information on the value of the EPC in a weaker fashion with respect to the intensities, because they scale as the integral of the square of the EPC even over non-resonant momenta.
Therefore, we plot the $A_{2D}/A_{2D'}$ ratio in Fig. \ref{fig2}(a) as a function of the laser excitation energy and of the position of the phonon resonance along the $\boldsymbol{\Gamma}-\mathbf{K}$ line (corresponding to the \textit{inner} processes of Ref. \cite{berciaud2013intrinsic}). We observe a huge (fivefold) enhancement of the experimental $A_{2D}/A_{2D'}$ ratio at 1.16 eV excitation energy for the graphene samples with respect to graphite. The intensity of the 2D and 2D$^\prime$ peaks of graphene has been also benchmarked against the intensity of the same resonant peaks for graphite (see Supplemental Material (SM) \cite{graphene2022_IR_raman}), showing that a difference between graphene and graphite is found at 1.16 eV for the 2D peak only and not for the 2D' peak. This qualitatively indicates an increase of the EPC in graphene at the $\mathbf{K}$ point with respect to graphite, while it remains comparable at the $\boldsymbol{\Gamma}$ point. Notably, the increase of the $A_{2D}/A_{2D'}$ ratio is found in graphene for both the encapsulated and free-standing samples, indicating that the dielectric screening of the environment is not playing a key role in this enhancement~\cite{forster2013dielectric}.

To provide a quantitative assessment of the modification of the dependence of the EPC on the vicinity to $\mathbf{K}$, we compare the experimental results with theoretical calculations.
The Raman spectrum $S(\omega)$ is calculated with the methodology of Refs. \cite{venezuela2011theory,PhysRevLett.113.187401}, as discussed in the SM. %
Here we only note that the energy-conserving Dirac-delta function is replaced with a Lorentzian-function that takes into account the lifetime of two phonons via $\gamma^{ph}=5.3$ cm$^{-1}$ \cite{paulatto2013anharmonic}. For the electron-hole broadening $\gamma^{ep}$ we consider only the interaction of carriers with phonons, and we use the expression deduced in Ref.~\cite{venezuela2011theory} for the conical model in the hypothesis of EPCs independent of the phonon momentum (see SM) as
\begin{align}
\gamma^{ep}(\omega_L)=82.7(\hbar\omega_L/2 - 0.166) \textrm{ meV},
\label{eq:gamma}
\end{align}
where $\hbar\omega_L$ is expressed in eV. 
The theoretical ratio $A_{2D}/A_{2D'}$ results to be numerically almost independent of the electron-hole linewidth $\gamma^{ep}$, in agreement with analytical calculations on the conical model (within a $\sim$15\% accuracy - see SM). Therefore, even from accurate \textit{ab-initio} calculations, it results that  $A_{2D}/A_{2D'}$ depends on the EPC only, at variance with the ratio of each single intensity with the G peak \cite{canccado2007measuring,PhysRevB.87.205435}.

The ratio $A_{2D}/A_{2D'}$ obtained via DFT calculations is depicted as a black dashed line in Fig.~\ref{fig2}(a): it is evident that calculations strongly underestimate the experimental results for visible excitation energies and, most importantly, are not able to catch the enhancement at low laser energies. The discrepancy between the theoretical calculation and the experimental data for phonons at zone boundary could be rooted in the neglect of the Coulomb vertex corrections to the dressing of the EPC \cite{basko2008interplay} proper of DFT in the common approximations for the electronic exchange-correlation term, e.g. in local density approximation~\cite{PhysRevLett.45.566,PhysRevB.23.5048} or generalized gradient approximations~\cite{PhysRevLett.77.3865}, as discussed in Ref. \cite{lazzeri2008impact}. As already mentioned, such corrections (of which the dominant contribution is shown in the inset of Fig. \ref{fig2}(b)) could be large for phonons near the zone-boundary $\mathbf{K}$, while are negligible at the Brillouin zone-center $\boldsymbol{\Gamma}$ \cite{Attaccalite2010,basko2008interplay}. 

To model the increase of the EPC at zone-boundary with respect to zone-center we introduce in the Kohn anomaly region at \textbf{K} a scaling parameter $r_{vc}(\omega_L) \equiv\langle D_{\mathbf{K}}^2\rangle/2\langle D_{\boldsymbol{\Gamma}}^2\rangle$, which takes into account the renormalization of the coupling due to the Coulomb vertex correction. Notice that in DFT $r_{vc}$ reduces to unity \cite{PhysRevLett.93.185503}. 
The ratio of the calculated peak intensities rescales with $r_{vc}$ via $A_{2D}/A_{2D'} \rightarrow r^2_{vc}A_{2D}/A_{2D'}$. We first rescale the theoretical results obtained by a constant value, $r_{vc}=1.4$, which is obtained from GW calculations for the vertex corrections to the EPC in graphite, and therein experimentally confirmed \cite{lazzeri2008impact, PhysRevB.80.085423}. With this rescaling we obtain the black dotted line shown in Fig.~\ref{fig2}, which is in qualitative agreement with the experimental points for graphite for all laser energies, and for graphene if $\hbar\omega_L > 2.33$ eV. The agreement between theory and experiment in graphite, even at $1.16\,$eV, shows that in the three-dimensional case the vertex corrections are weakly dependent on the distance from $\mathbf{K}$. On the other hand, the curve clearly fails in reproducing the huge enhancement found experimentally for graphene for $\hbar\omega_L < 2.33$ eV. In this case, $r_{vc}$ can therefore not be taken as a constant independently of the laser energy, i.e. independently of the distance from the $\mathbf{K}$ point. By the comparison with the experimental data we find $r_{vc}$ to be
\begin{equation}
r_{vc}(\omega_L)=
\begin{cases}
1.4 \qquad \hbar\omega_L > 2.33\, \textrm{eV} \quad |\mathbf{q}-\mathbf{K}|> 0.39\;\angstrom^{-1}
\\
4.8 \qquad \hbar\omega_L = 1.16\, \textrm{eV} \quad |\mathbf{q}-\mathbf{K}|= 0.16\;\angstrom^{-1}
\end{cases},
\label{eq:r_cv}
\end{equation}
where we indicated also the corresponding position of the phonon resonance along the $\boldsymbol{\Gamma}-\mathbf{K}$ line.
For intermediate points or for laser energies lower than 1.16 eV we fit a second-order polynomial law for $r_{vc}(\omega_L)$ to our experimental data (see SM); the rescaled curve for $A_{2D}/A_{2D'}$ is presented in Fig. \ref{fig2}(a) as a solid line. Finally, to quantify the effect of this rescaling on the EPC, we plot in Fig. \ref{fig2}(b) the change of $\langle D^2_{\mathbf{K}}\rangle$ due to the modification of the resonance condition as a function of the distance of the resonant phonon momentum \textbf{q} from the \textbf{K} point, as deduced from the expression of $r_{vc}(\omega_L)$ by taking the value of $\langle D^2_{\boldsymbol{\Gamma}}\rangle$ from GW calculations on graphite \cite{lazzeri2008impact}, and supposing that $\langle D^2_{\boldsymbol{\Gamma}}\rangle$ does not critically depend on the dimensionality. At $1.16\,$eV we find that $\langle D^2_{\mathbf{K}}\rangle^\mathrm{graphene} \sim 560 $ (eV/ $\textup{\AA})^2$ as
$\langle D^2_{\mathbf{K}}\rangle^\mathrm{graphene} / \langle D^2_{\mathbf{K}}\rangle^\mathrm{graphite} \sim 3.4$. Since the EPC is expected to increase for $|\mathbf{q}-\mathbf{K}| \rightarrow 0$, the value of $\langle D^2_{\mathbf{K}}\rangle$ at $0.16\;$\AA$^{-1}$ can be thus used as the lower bound for the electron-phonon interaction at \textbf{K}.
\begin{figure} [ht!]
\begin{center}
\includegraphics[scale=0.38]{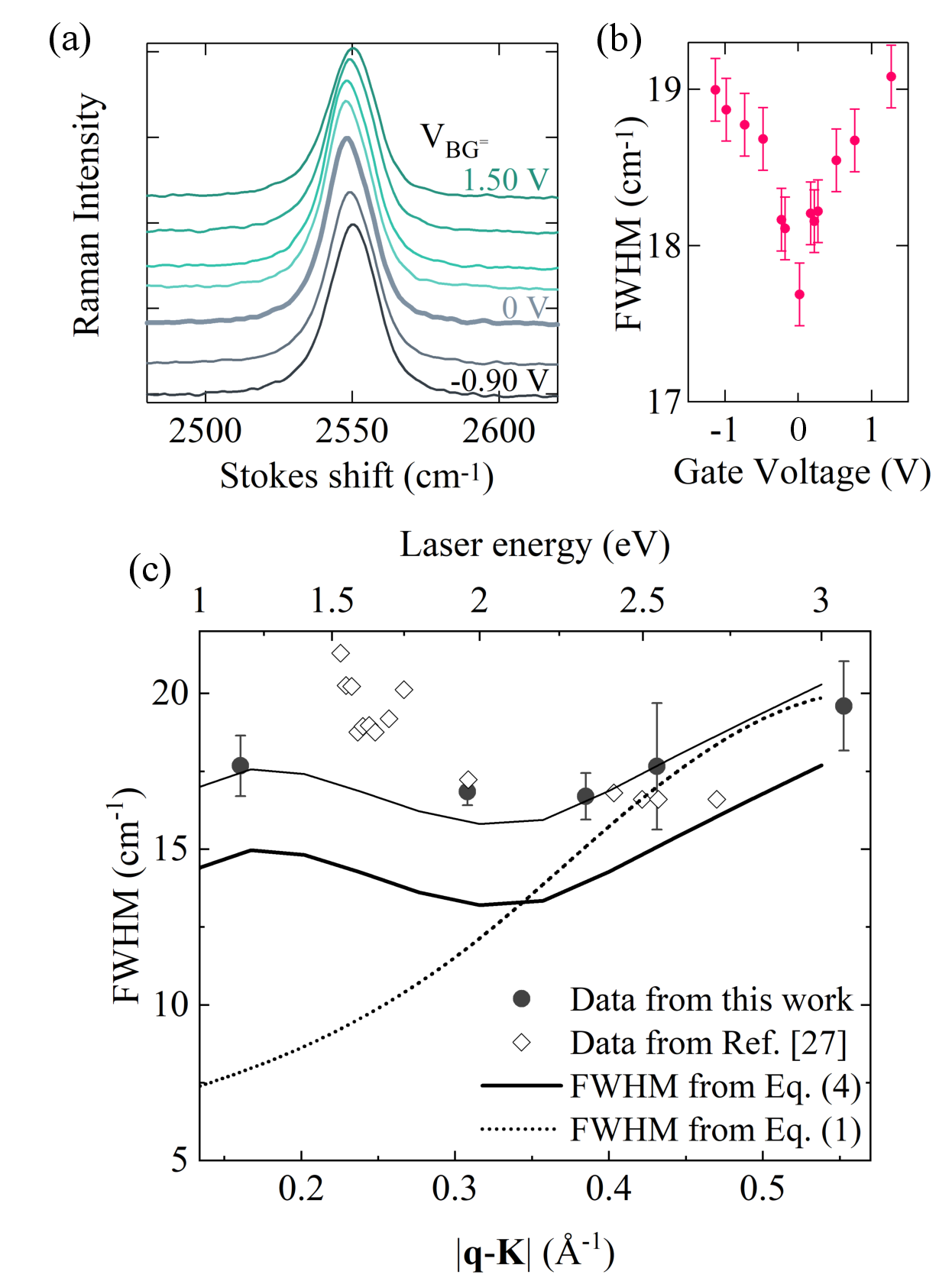}
\caption{(a) Raman spectra at different gate voltage values from which the FWHM of the 2D peak is extracted. We note that indeed the FWHM is minimum for gate voltages close to 0 V, indicating that our unbiased sample is undoped. (b) FWHM measured at 1.16 eV as a function of back gate voltage. (c) FWHM as a function of laser excitation energy (full markers), extracted numerically from the fitted data, compared to data from Ref.~\cite{berciaud2013intrinsic} (empty markers). The experimental data are compared to the result of the theoretical calculation (dotted curve) with EPC constant assumed independent of the excitation laser energy. The solid curve is obtained instead using Eq.~\ref{eq:newgamma}, i.e.\ by rescaling the EPC at \textbf{K} with $r_{vc}(\omega_L)$. We have also rigidly shifted the latter curve by 2 cm$^{-1}$ to better highlight how it compares to the experimental data (thinner solid line).}
\label{fig3} 
\end{center}      
\end{figure}

To corroborate our analysis with other experimental evidences, we study how $r_{vc}(\omega_L)$ impacts the shape of the Full Width at Half Maximum (FWHM) of the 2D peak. Indeed, discrepancies between the experimental data and the theoretical FWHM were already seen in Ref.~\cite{berciaud2013intrinsic}. Therein, by decreasing $\omega_L$ from the visible towards the infrared, the FWHM does not decrease by following the expected trend, that is given by ~\cite{basko2008theory,venezuela2011theory}:
\begin{equation}
\FWHM_{2D}(\omega_L) = 2\sqrt{2^{2/3} -1} \frac{v_{ph}}{v_F}\gamma^{ep}(\omega_L)+\gamma^{ph},
\label{eq:FWHM}
\end{equation}
where $v_{ph}/v_F$ is the phonon/Fermi velocity at a given laser energy $\omega_L$ and $\gamma^{ep}(\omega_L)$ is given by Eq. (\ref{eq:gamma}).

In order to extract the FWHM of the 2D peak in the condition of zero doping, i.e. minimum linewidth ~\cite{sonntag2023charge}, we have measured the Raman spectra as a function of gate voltage V$_{G}$ (Fig.\ref{fig3} (a)), verifying that for V$_{G}$= 0 we are close to zero doping (Fig.\ref{fig3} (b)). Furthermore, we have measured a Raman map to perform the measurements in a homogeneous region of the graphene flake in such a way to minimize the inhomogenous broadening of the 2D peak (see section S6 of the SM). In Fig.~\ref{fig3}(c) we display the experimental FWHM of the 2D peak (data points) extracted by fitting the data with the sum of two Baskovian functions (see SM) and by numerically taking the FWHM from the best-fit function. We notice that the FWHM of the 2D peak does not decrease with decreasing $\hbar\omega_L$, as opposed to the result of \textit{ab initio} calculations (dotted black curve) that decreases with $\hbar\omega_L$ similarly to the analytical calculation on the conical model given by Eq. \ref{eq:FWHM} (see SM). To show that this discrepancy is due to the increase of $r_{vc}$ at lower laser frequencies, we modify the excitation energy dependency of $\gamma^{ep}$ in Eq. \ref{eq:FWHM} by supposing that the value of $\langle D_{\boldsymbol{\Gamma}}^2\rangle$ for graphene does not depend on the vicinity to $\boldsymbol{\Gamma}$, while the value of $\langle D_{\mathbf{K}}^2\rangle$ has to be enhanced via the use of $r_{vc}(\omega_L)$. We eventually obtain (see SM):
\begin{align}
\gamma^{ep}=\left[\frac{\hbar\omega_L}{2}(34.3r_{vc}(\omega_{L})+26.4)-5.2\big(r_{vc}(\omega_L)+1\big)\right] \textrm{meV}.
\label{eq:newgamma}
\end{align}
Eq. (\ref{eq:newgamma}) yields an almost constant behaviour of the FWHM as a function of the laser excitation energy (solid curve in Fig.~\ref{fig3}(c)), resembling the trend of the experimental data, only with a slight offset. 

In conclusion, we report a comprehensive resonance Raman spectroscopy experiment on graphene with laser excitation energy down to the infrared ($\hbar\omega_L$= 1.16 eV). We find a huge enhancement of the intensity ratio of the resonant 2D and 2D$^\prime$ peaks $A_{2D}/A_{2D'}$ that we explain with an enhancement of the electron-phonon coupling while approaching $\bf{K}$, excluding a crucial role of the dielectric environment on the phenomena observed. This finding applies to all 2D materials where Coulomb interaction could induce a strong enhancement of the EPC in the neighborhood of the Dirac point and could moreover have a significant impact on the modeling of transport in graphene devices at RT where optical phonons contribute to the resistivity comparably to the acoustic modes~\cite{park2014electron}. By developing a resonance Raman setup with even smaller excitation energy, i.e. probing the linear electronic dispersion as close as possible to the Dirac point, one will be able to directly address electron-phonon and electron-electron interactions in all sorts of graphene devices and in those systems where the linear electronic dispersion occurs only in small energy ranges.

\begin{acknowledgments}
We acknowledge the European Union's Horizon 2020 research and innovation program under grant agreements no. 881603-Graphene Core3 and the MORE-TEM ERC-SYN project, grant agreement No 951215. We acknowledge that the results of this research have been achieved using the DECI resource \textit{Mahti CSC} based in Finland at https://research.csc.fi/-/mahti with support from the PRACE aisbl; we also acknowledge PRACE for awarding us access to Joliot-Curie Rome at TGCC, France. We acknowledge funding from Sapienza University of Rome through the \textit{Medie Attrezzature} program. We acknowledge support from the PRIN2017 Grant No. 2017Z8TS5B. T. O., B. B. and C. S. acknowledge funding from the Deutsche Forschungsgemeinschaft (DFG, German Research Foundation) under Germany’s Excellence Strategy - Cluster of Excellence Matter and Light for Quantum Computing (ML4Q) EXC 2004/1–390534769. Co-funded by the European Union (ERC, DELIGHT, 101052708 and
NextGenerationEU). Views and opinions expressed are however those of the
author(s) only and do not necessarily reflect those of the European
Union or the European Research Council. Neither the European Union nor
the granting authority can be held responsible for them.
\end{acknowledgments}

\section*{}
\noindent T.V. and L.G contributed equally to this work.


\end{document}